\documentstyle[epsf]{mn2e}

\title[On the wake generated by a planet in a disc]
{On the wake generated by a planet in a disc}

\author[G. I. Ogilvie and S. H. Lubow]
  {G. I. Ogilvie$^1$ and S. H. Lubow$^{1,2}$\\
  $^1$Institute of Astronomy, University of Cambridge, Madingley Road,
  Cambridge CB3 0HA\\
  $^2$Space Telescope Science Institute, 3700 San Martin Drive,
  Baltimore, MD 21218, USA}

\date{Accepted 2001 November 12.  Submitted 2001 August 14}

\begin{document}

\maketitle

\label{firstpage}

\begin{abstract}
  A planet of low mass orbiting in a two-dimensional gaseous disc
  generates a one-armed spiral wake.  We explain this phenomenon as
  the result of constructive interference between wave modes in the
  disc, somewhat similar to the Kelvin wedge produced in the wake of a
  ship.  The same feature is not expected in a three-dimensional disc
  with thermal stratification.
\end{abstract}


\section{Introduction}

The dynamical interaction between a planet and the disc in which it
forms has important consequences for the evolution of the orbital
elements of the planet.  In particular, any imbalance between the
torques exerted on the planet by the parts of the disc interior and
exterior to its orbit causes the planet to migrate radially through
the disc (e.g. Ward 1986).  The planet--disc interaction has been
studied using linear theory and, more recently, non-linear numerical
simulations.  Most of this work uses a two-dimensional description
that ignores the vertical structure and motion of the disc.

In the linear theory (Goldreich \& Tremaine 1979) the planetary
potential is analysed into Fourier components, each having a definite
azimuthal wavenumber and angular pattern speed.  For each Fourier
mode, the response of the disc is determined by an inhomogeneous
linear wave equation that includes resonances at various radii.
Angular momentum is exchanged between the planet and the disc mainly
in the vicinity of Lindblad resonances (at which waves are launched)
and the corotation resonance (where angular momentum is deposited).
Goldreich \& Tremaine (1979) provided analytical formulae for these
torques, while Korycansky \& Pollack (1993) determined them by solving
the linear wave equations numerically.

Neither of these studies was concerned with the form of the
disturbance generated in the disc, which is obscured in the Fourier
representation.  In contrast, numerical simulations of the
planet--disc interaction (e.g. Bryden et al. 1999; Kley 1999; Lubow,
Seibert \& Artymowicz 1999) aim to solve the non-linear fluid
dynamical equations in real space.  Most such simulations have treated
the case of a Jovian planet, for which the planet--disc interaction is
highly non-linear, especially close to the planet where a gap is
opened in the disc.  Further away, spiral shock waves are distinctly
seen.

A detailed comparison between linear and non-linear theory has been
made by Miyoshi et al. (1999), who compared the torques generated in a
local, three-dimensional simulation (an isothermal, shearing-sheet
model) with those derived by solving the linear wave equations.  They
did not compare the form of the disturbance in real space, however.

For planets of lower mass, a gap is not opened and the response of the
disc should be well described by linear theory.  Simulations for a
terrestrial planet (Artymowicz 2001) indicate that a remarkable
one-armed spiral wake is formed, but there is no evidence for features
associated with individual Lindblad resonances.  The relation between
the wake and the linear Fourier analysis has been obscure.  In this
letter we explain this relation by showing how the wake is formed from
the superposition of individual Fourier modes.  We also discuss the
differences that are to be expected in future three-dimensional
simulations.

\section{Linear hydrodynamic waves in a two-dimensional disc}

\subsection{Dispersion relation and pitch angle}

Consider a two-dimensional gaseous disc around a star of mass $M$.
Let the disc have angular velocity $\Omega(r)$, surface density
$\Sigma(r)$ and adiabatic sound speed $v_{\rm s}(r)$.  Suppose that a
planet of mass $M_{\rm p}$ orbits within the disc on a circular orbit
of radius $r_{\rm p}$ and angular velocity $\Omega_{\rm p}$.  The
planet interacts gravitationally with the disc and excites
hydrodynamic waves having angular pattern speed $\Omega_{\rm p}$.  We
assume that the waves may be described using linear theory.

For linear waves, any wave quantity $X$ can be written in the form
\begin{equation}
  X(r,\phi,t)={\rm Re}\left\{\tilde X(r)
  \exp\left[{\rm i}\Phi_m(r,\phi,t)\right]\right\},
\end{equation}
where $\tilde X$ is an amplitude that varies slowly with $r$, while
\begin{equation}
  \Phi_m=\int k(r)\,{\rm d}r+m(\phi-\Omega_{\rm p}t)
  \label{Phi_m}
\end{equation}
is a phase that varies rapidly with $r$.  Here $k$ is the radial
wavenumber, which is real in regions of space where the wave
propagates, and $m$ is the azimuthal wavenumber.  We consider only
non-axisymmetric waves, and take $m$ to be a positive integer by
convention.  We initially leave the constant of integration in
equation (\ref{Phi_m}) indefinite.

The dispersion relation for tightly wound hydrodynamic waves in
a two-dimensional gaseous disc is
\begin{equation}
  \left[m(\Omega-\Omega_{\rm p})\right]^2=\kappa^2+v_{\rm s}^2k^2,
  \label{dispersion}
\end{equation}
where $\kappa(r)$ is the epicyclic frequency.  We neglect the effects
of self-gravitation and dissipative processes.  The wavefronts, or
lines of constant phase, are spirals defined by the equation
\begin{equation}
  {{{\rm d}\phi}\over{{\rm d}r}}=-{{k}\over{m}}.
\end{equation}
We note that, if the term $\kappa^2$ in equation (\ref{dispersion})
may be neglected compared to the other terms, the pitch angle
$\arctan(m/kr)$ is independent of $m$.  This raises the possibility,
in principle, that constructive interference may occur between waves
of different azimuthal wavenumber.  Since $\kappa$ and $\Omega$ are
usually of comparable magnitude (indeed, they are equal for a
Keplerian disc), the term $\kappa^2$ is relatively small when $m$ is
large, except close to the corotation radius where $\Omega=\Omega_{\rm
  p}$.

\subsection{Disc model}

\label{Disc model}

We assume that the disc is Keplerian, so that\footnote{Although torque
  calculations are sensitive to slight deviations from Keplerian
  rotation, e.g. resulting from radial pressure gradients, they are
  not important in the present analysis.}
\begin{equation}
  \kappa=\Omega=\left({{GM}\over{r^3}}\right)^{1/2}.
\end{equation}
The corotation radius is then
\begin{equation}
  r_{\rm c}=\left({{GM}\over{\Omega_{\rm p}^2}}\right)^{1/3}.
\end{equation}
We assume further that the sound speed is given by
\begin{equation}
  v_{\rm s}=\epsilon\left({{GM}\over{r}}\right)^{1/2},
  \label{v_s}
\end{equation}
where $\epsilon$ is a constant.  This convenient assumption, which is
not critical to the analysis that follows, implies that the disc has a
constant angular semi-thickness $H/r$ proportional to $\epsilon$, and
a constant Mach number $\epsilon^{-1}$.

Adopting units such that $GM=\Omega_{\rm p}=r_{\rm c}=1$, we find
\begin{equation}
  k^2={{m^2}\over{\epsilon^2r^2}}
  \left(r^{3/2}-r_+^{3/2}\right)\left(r^{3/2}-r_-^{3/2}\right),
\end{equation}
where
\begin{equation}
  r_\pm=\left(1\pm{{1}\over{m}}\right)^{2/3}
\end{equation}
are the radii of the outer and inner Lindblad resonances for mode $m$.
Waves are launched by the planet at the Lindblad resonances and
propagate into $r>r_+$ (with $k>0$) and into $r<r_-$ (also with
$k>0$).  The sign of $k$ is chosen such that the radial group velocity
is directed away from the planet.

\subsection{Waves in the outer disc}

Consider the waves launched in the outer disc, i.e. the disc exterior
to the planet's orbit.  The precise phase of mode $m$ for $r>r_+$ is
given by
\begin{equation}
  \Phi_m={{\pi}\over{4}}+\int_{r_+}^rk(r')\,{\rm d}r'+m(\phi-t),
\end{equation}
where the term $\pi/4$ is the phase shift associated with the
resonance.  This can be deduced from the behaviour of the Airy
function, which describes the launched wave near the Lindblad
resonance.

We first obtain an {\it estimate\/} of the phase by a method
equivalent to neglecting the term $\kappa^2$ in the dispersion
relation.  As noted above, this is appropriate when $m$ is large.
Approximating $r_\pm\approx1$, we have
\begin{equation}
  \int k(r)\,{\rm d}r\approx{{m}\over{\epsilon}}
  \int r^{-1}\left(r^{3/2}-1\right)\,{\rm d}r,
\end{equation}
and therefore
\begin{equation}
  \Phi_m\approx{{\pi}\over{4}}+
  {{2m}\over{3\epsilon}}\left(r^{3/2}-{{3}\over{2}}\ln r-1\right)+
  m(\phi-t).
\end{equation}
Constructive interference occurs near the curve $\phi=\varphi(r,t)$
defined by
\begin{equation}
\label{phaseouter}
  \varphi=t-{{2}\over{3\epsilon}}\left(r^{3/2}-{{3}\over{2}}\ln r-1\right),
\end{equation}
since we have
\begin{equation}
  \Phi_m\approx{{\pi}\over{4}}+m(\phi-\varphi).
\end{equation}

We can, however, calculate the phase exactly by changing to the
variable $x=r^{3/2}$ and making use of the indefinite integral
\begin{eqnarray}
  \lefteqn{\int x^{-1}(x-a)^{1/2}(x-b)^{1/2}\,{\rm d}x=
  (x-a)^{1/2}(x-b)^{1/2}}&\nonumber\\
  &&-(a+b)\ln\left[(x-a)^{1/2}+(x-b)^{1/2}\right]-(ab)^{1/2}\nonumber\\
  &&\times\ln\left[{{(a+b)x-2ab-2(ab)^{1/2}(x-a)^{1/2}(x-b)^{1/2}}
  \over{x}}\right].\nonumber\\
\end{eqnarray}
We write the exact solution as
\begin{equation}
  \Phi_m={{\pi}\over{4}}+m(\phi-\varphi)-{{2}\over{3\epsilon}}\Delta_m(r).
\end{equation}
The last term measures the error in the approximate method, and will
determine whether constructive interference really does occur on the
curve $\phi=\varphi$.  The approximate method always slightly
overestimates both the wavenumber $k$ and the range of integration in
the phase integral.  It follows that the residue $\Delta_m$ has the
properties
\begin{equation}
  \Delta_m>0,\qquad
  {{{\rm d}\Delta_m}\over{{\rm d}r}}>0.
  \label{monotonic}
\end{equation}
The expansion for large $r$ is
\begin{equation}
  \Delta_m(r)=\Delta_m(\infty)-{{1}\over{2mr^{3/2}}}-
  {{1}\over{4mr^3}}+O(r^{-9/2}),
\end{equation}
where (see Fig.~1)
\begin{equation}
  \Delta_m(\infty)=m\ln(2m)+
  (m^2-1)^{1/2}\ln\left[m-(m^2-1)^{1/2}\right].
\end{equation}

\begin{figure}
  \centerline{\epsfbox{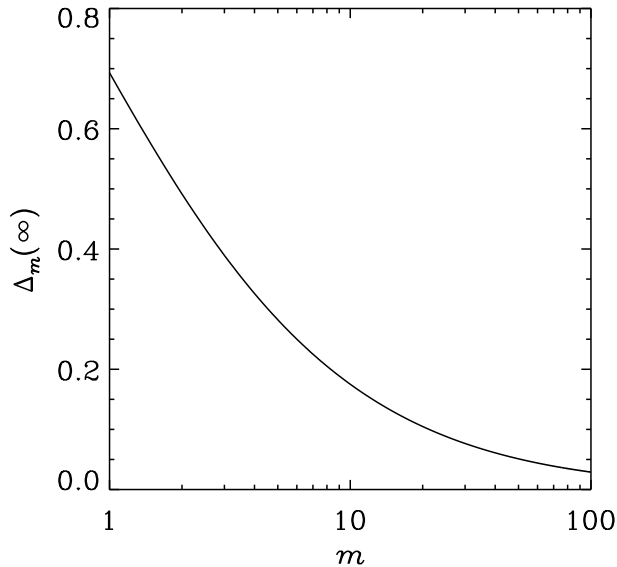}}
  \caption{The residue $\Delta_m(\infty)$.}
\end{figure}

A further property is that, for any fixed $r$,
\begin{equation}
  \lim_{m\to\infty}\Delta_m=0.
  \label{limit}
\end{equation}
In fact, for large $m$,
\begin{equation}
  \Delta_m\sim{{1}\over{2m}}
  \ln\left[2{\rm e}^{1/2}m\left(1-r^{-3/2}\right)\right].
\end{equation}

Waves of different $m$ add constructively on the curve $\phi=\varphi$
provided that their phases lie within a range less than approximately
$\pi$.  This certainly occurs for waves of sufficiently large $m$,
because of property (\ref{limit}).  Constructive interference may fail
for low values of $m$ for which $\Delta_m\ga3\pi\epsilon/2$, since
such modes are out of phase with waves of high $m$.  The properties
(\ref{monotonic}) ensure that the limit $r\to\infty$ is the worst
case.  The residue is plotted in Fig.~2 for $m=1,\dots,10$.  For
$\epsilon=0.1$, typical of protoplanetary discs, constructive
interference fails only for $1\le m\le2$ in the limit of large $r$.

\begin{figure}
  \centerline{\epsfbox{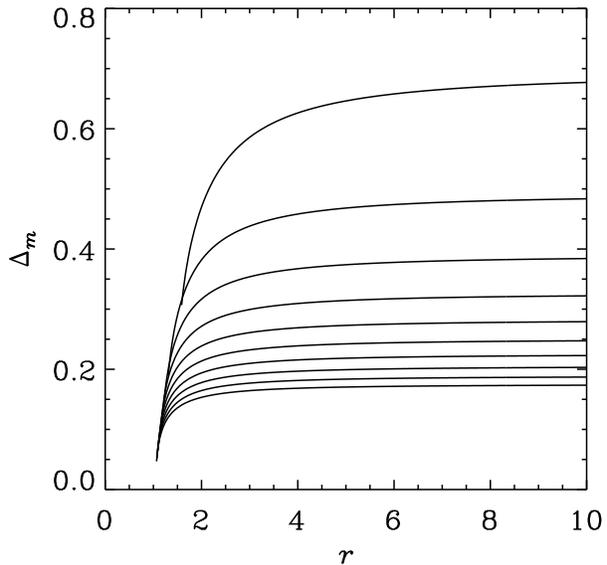}}
  \caption{The residue $\Delta_m$ in the outer disc.  Modes
    $m=1,\dots,10$ are shown from top to bottom.}
\end{figure}

Provided that a range of azimuthal wavenumbers is present, and there
is no special selection rule (e.g. one that selects azimuthal
wavenumbers that are all multiples of $2$), there is no other curve on
which constructive interference occurs consistently.  The result is a
{\it one-armed spiral wake\/} following the curve $\phi=\varphi$.

The disturbance generated by a planet orbiting in the disc is
dominated by azimuthal wavenumbers $m\approx m_*\approx1/(2\epsilon)$
(Goldreich \& Tremaine 1980).  Accordingly, the azimuthal thickness of
the wake is $\delta\phi\approx2\pi/m_*\approx4\pi\epsilon$.  The
radial thickness is smaller, $\delta
r/r\approx\epsilon\,\delta\phi\approx4\pi\epsilon^2$.

\subsection{Waves in the inner disc}

\begin{figure}
  \centerline{\epsfbox{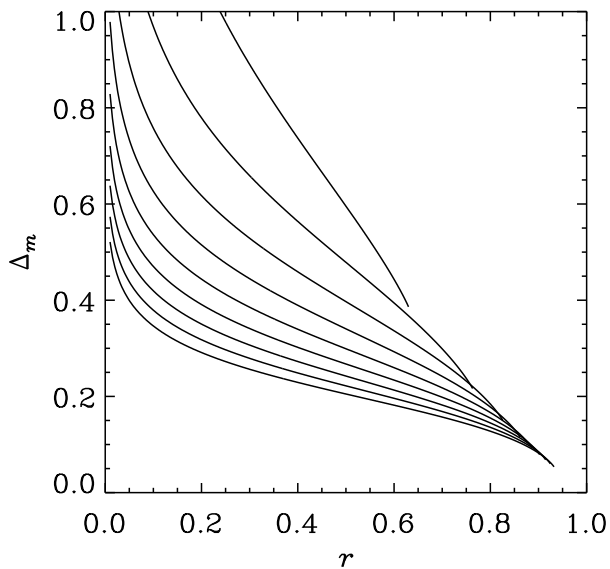}}
  \caption{The residue $\Delta_m$ in the inner disc.  Modes
    $m=1,\dots,10$ are shown from top to bottom.}
\end{figure}

Waves launched in the inner disc behave very similarly.  The precise
phase of mode $m$ for $r<r_-$ is given by
\begin{equation}
  \Phi_m={{\pi}\over{4}}+\int_{r_-}^rk(r')\,{\rm d}r'+m(\phi-t),
\end{equation}
where, again, $k>0$.
We again write the exact solution in the form
\begin{equation}
  \Phi_m={{\pi}\over{4}}+m(\phi-\varphi)+{{2}\over{3\epsilon}}\Delta_m(r),
\end{equation}
where now
\begin{equation}
\label{phaseinner}
  \varphi=t+{{2}\over{3\epsilon}}\left(r^{3/2}-{{3}\over{2}}\ln r-1\right).
\end{equation}
In the inner disc the residue has the properties
\begin{equation}
  \Delta_m>0,\qquad
  {{{\rm d}\Delta_m}\over{{\rm d}r}}<0.
\end{equation}
The limiting form for small $r$ is
\begin{equation}
  \Delta_m(r)=-{{3}\over{2}}\left[m-(m^2-1)^{1/2}\right]\ln r+O(1).
\end{equation}
Constructive interference does {\it eventually\/} fail for all $m$ in
the limit $r\to0$.

The residue is plotted in Fig.~3 for $m=1,\dots,10$.  The limiting
form for large $m$ is
\begin{equation}
  \Delta_m\sim{{1}\over{2m}}
  \ln\left[2{\rm e}^{1/2}m\left(r^{-3/2}-1\right)\right].
\end{equation}

\section{Numerical calculation}

\label{Numerical calculation}

To verify the hypothesis that the wake is formed through the
constructive interference of Fourier modes, we have calculated the
linear response of the disc numerically using methods similar to
Korycansky \& Pollack (1993).\footnote{We caution the reader of that
  paper of a number of errors, especially on p.~164.}  We
adopted a two-dimensional barotropic disc model equivalent to that
described in Section~\ref{Disc model}, and having a surface density
$\Sigma\propto r^{-3/2}$.  Outgoing-wave boundary conditions were
applied at $r=0.3$ and $r=3$.  The potential of the planet was
smoothed to simulate the non-zero vertical extent of the disc, with a
smoothing length $\epsilon r_{\rm p}$ comparable to the semi-thickness
$H$.  The linear solutions for modes $m=1$ to $m=100$ were calculated
and synthetized in real space.

The results are shown in Figs~4 and~5, where we compare the predicted
shape of the wake given by equations (\ref{phaseouter}) and
(\ref{phaseinner}) with the numerical calculation.  The agreement is
excellent.  The numerical calculation shows that the wake has some
non-trivial internal structure, often consisting of both a trough and
a larger peak.  These details are likely to depend to some extent on
the disc model.

\begin{figure*}
  \centerline{\epsfysize=8cm\epsfbox{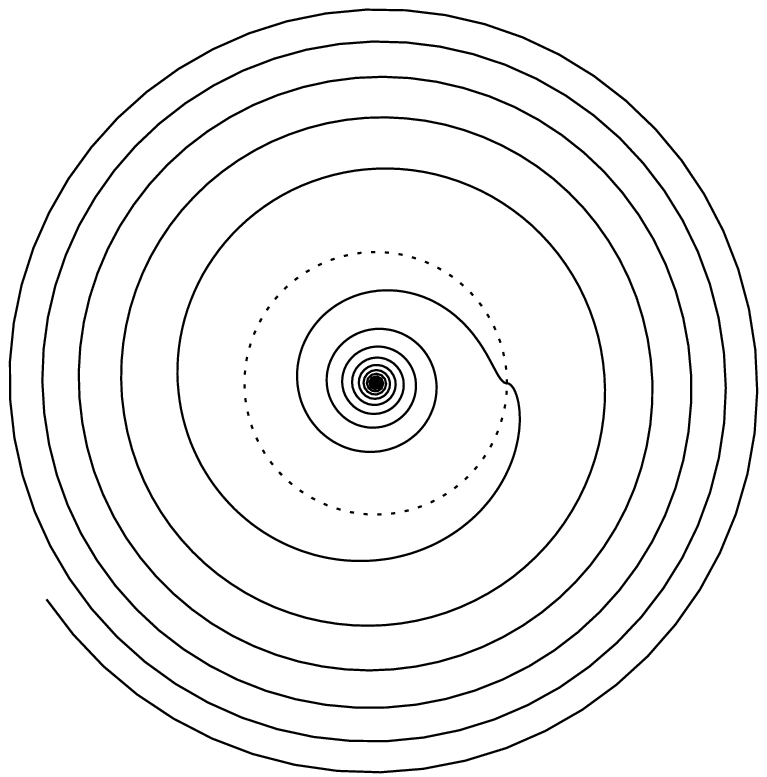}\qquad\epsfysize=8cm\epsfbox{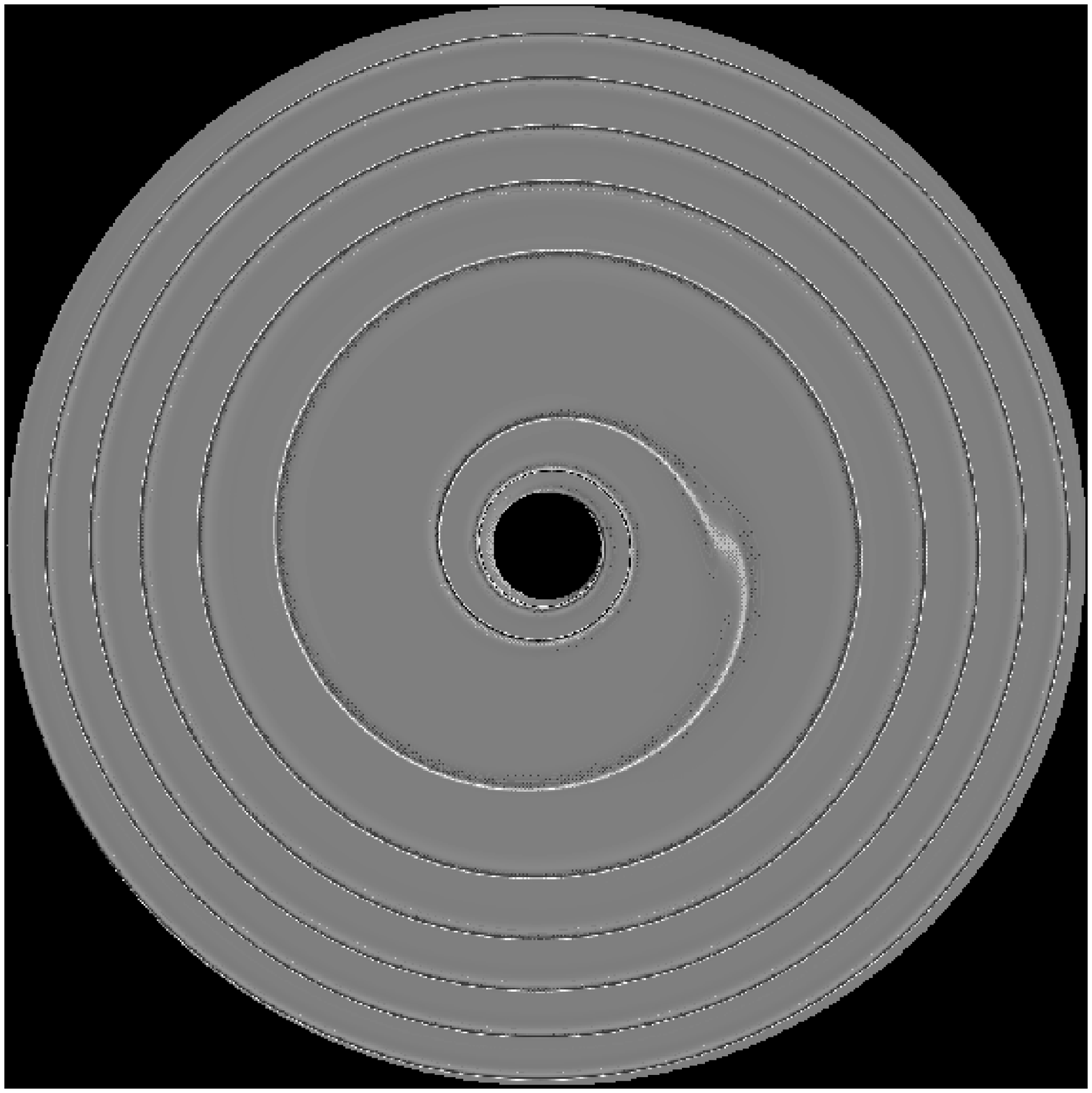}}
  \caption{Left: Predicted shape of the spiral wake for $\epsilon=0.05$,
    based on equations (\ref{phaseouter}) and (\ref{phaseinner}).  The
    dotted line represents the corotation circle, $r=1$.  The planet
    is located at $(1,0)$, and the outer radius plotted is $r=3$.
    Right: Numerically calculated spiral wake for $\epsilon=0.05$.
    The enthalpy perturbation is plotted using a linear grey-scale
    from negative (black) to positive (white).  The maximum intensity
    corresponds to a fractional surface density perturbation, at
    $r=1$, of $10^4(M_{\rm p}/M)$.}
\end{figure*}

\begin{figure*}
  \centerline{\epsfysize=8cm\epsfbox{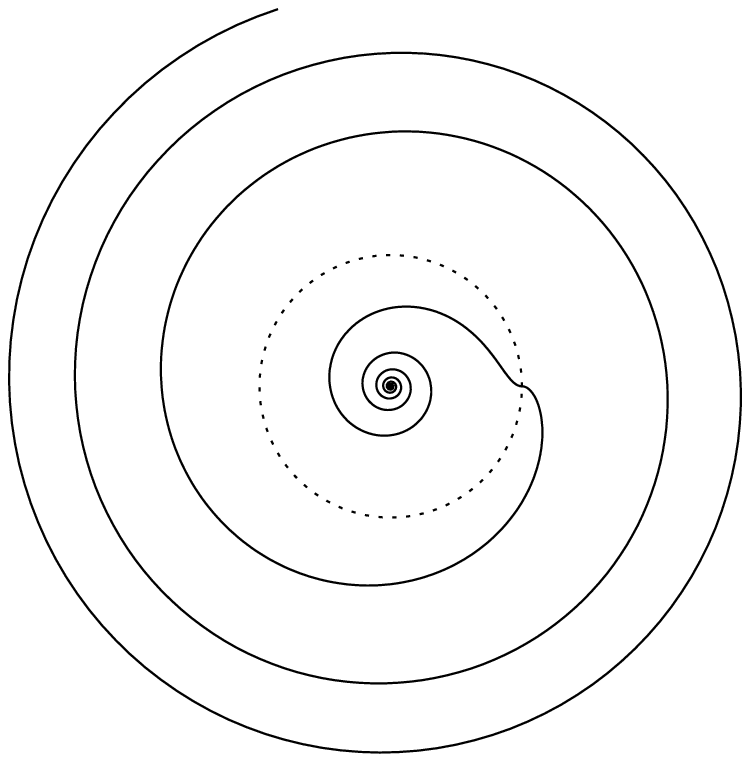}\qquad\epsfysize=8cm\epsfbox{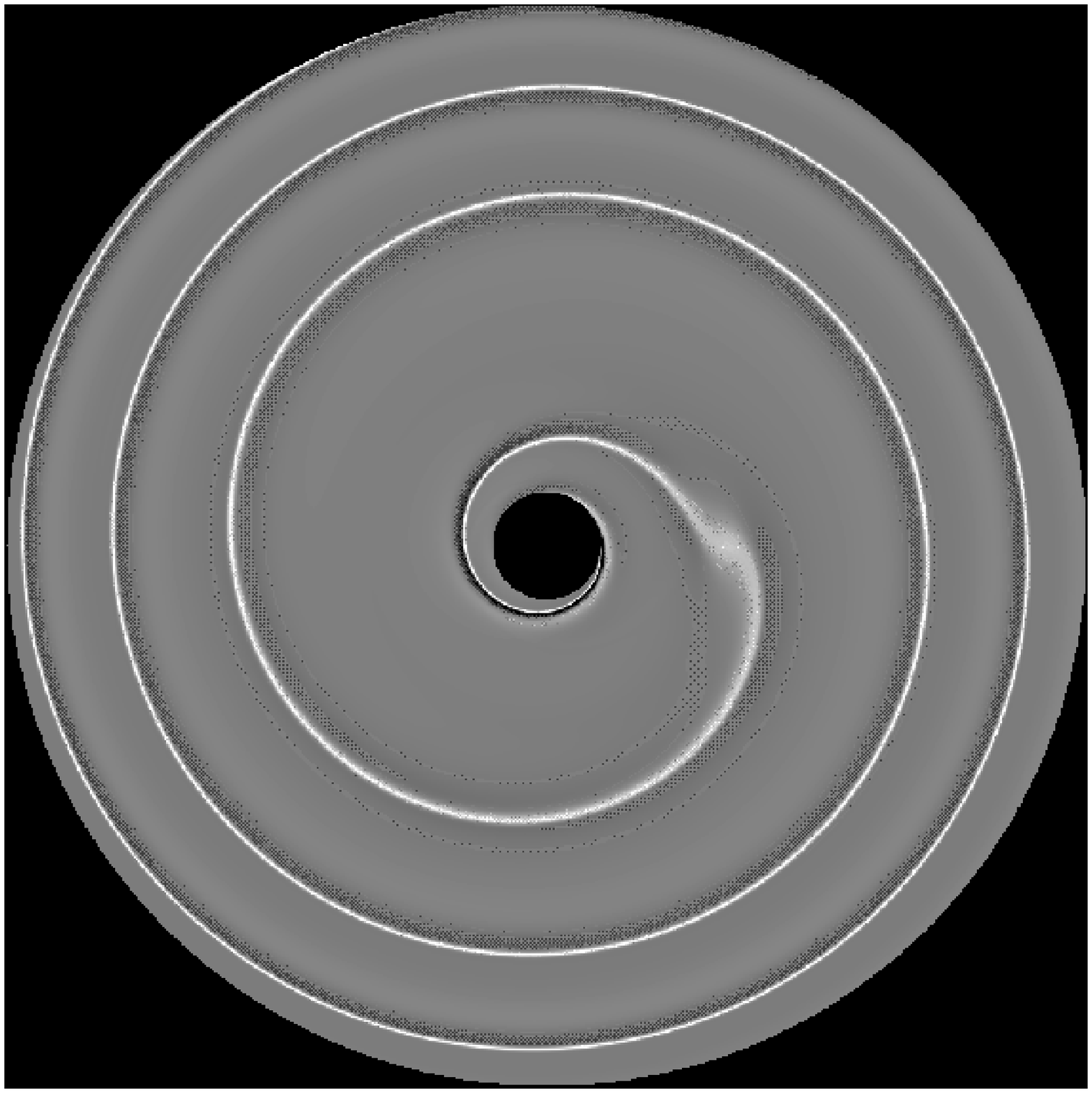}}
  \caption{As for Fig.~4, but with $\epsilon=0.1$.  The scale of the
    perturbation is $10^3(M_{\rm p}/M)$.}
\end{figure*}

Although the waves are generated at Lindblad resonances, individual
resonances cannot be observed in the complete solution because they
overlap.  The wake is generated in a perfectly smooth manner.  It is
clearly established after the synthesis of the first 10 modes in the
case of $\epsilon=0.05$, or the first 5 modes in the case of
$\epsilon=0.1$.  The addition of higher-order modes merely increases
the finesse of the interference pattern.

\section{Discussion}

We have considered the dynamical interaction between a planet of low
mass and a two-dimensional gaseous disc in which it orbits.  Although
the planet generates disturbances of all azimuthal wavenumbers, the
wave modes interfere constructively on a unique curve and form a
one-armed spiral wake.  The waveform based on linear theory (see
Figs~4 and~5) is in good agreement with that found in non-linear
planet--disc simulations (e.g. Artymowicz 2001).

The formation of a coherent structure by constructive interference is
reminiscent of the Kelvin wedge produced in the wake of a ship.  In
the present case the effect is more complete, as almost all modes
interfere constructively, rather than a band of wavenumbers.

Recently, Goodman \& Rafikov (2001) calculated the linear wake
produced by a terrestrial planet in a local, shearing-sheet model of a
two-dimensional disc.  As they noted, the formation of the wake
enhances the amplitude of the disturbance in a thin structure and
therefore increases the likelihood of non-linear effects such as shock
formation.

The precise relation between a linear wake and the spiral shocks
observed in numerical simulations involving planets of larger mass
(e.g. Lubow et al. 1999) is not entirely clear, however.  In such
simulations a dominant one-armed shock is formed, which follows a
similar curve to the linear prediction, but other features are also
present.  The flow near the planet is also different in the non-linear
regime.

Although most treatments of the planet--disc interaction have used a
two-dimensional description of the disc, this is in fact difficult to
justify.  First, the wave modes in a three-dimensional disc are, in
general, different from those in a two-dimensional disc.  If the disc
is vertically isothermal, there does exist a two-dimensional mode with
the same dispersion relation as equation (\ref{dispersion}) (Lubow \&
Pringle 1993).  In that case, a wake would be formed in much the same
way.  In a vertically thermally stratified disc, however, the waves
generated by the planet have a quite different dispersion relation.
For example, in a polytropic model, which represents a highly
optically thick disc with vertically distributed energy dissipation,
the equivalent mode behaves like a surface gravity wave having an
approximate dispersion relation
\begin{equation}
  \left[m(\Omega-\Omega_{\rm p})\right]^2\approx gk
\end{equation}
sufficiently far from the planet, where $g=\Omega^2H$ is the vertical
gravitational acceleration at the surface of the disc (Ogilvie 1998;
Lubow \& Ogilvie 1998).  Spiral waves with different values of $m$
then have different pitch angles and there is no possibility of
consistent constructive interference on a curve.\footnote{Although
  surface gravity waves are precisely what is involved in a ship wake,
  the analogy with the Kelvin wedge breaks down when the shear in the
  disc is taken into account.}  We therefore anticipate that the
disturbance generated by a planet in a thermally stratified disc will
have a quite different structure.

In addition, the flow near the planet is likely to be
three-dimensional if the radius of the planet's Roche lobe is
comparable to or less than the semi-thickness of the disc.  Therefore,
although this analysis may explain some features observed in numerical
simulations, the realities of the planet--disc interaction are likely
to be more complicated.

We remark that an accurate linear calculation of the planetary wake,
and of the associated migration rate, in a thermally stratified,
three-dimensional disc would be a demanding numerical problem, and has
not been carried out.

In summary, the one-armed wake is a consequence of having a spectrum
of two-dimensional waves with different azimuthal wavenumbers $m$ and
approximately matching phases.  Any mechanism capable of producing
such waves will result in a similar one-armed wake.  We have shown
that resonantly launched waves in a two-dimensional gaseous disc
naturally meet this criterion.

\section*{Acknowledgments}

We are grateful to Marcus Roper for carrying out an early version of
the calculation in Section~\ref{Numerical calculation}.  GIO
acknowledges the support of the Royal Society through a University
Research Fellowship.  SHL acknowledges support from the Institute of
Astronomy visitor programme and from NASA grants NAG5-4310 and
NAG5-10732.

\label{lastpage}

\end{document}